# Exact S-Wave Solution of the Trigonometric Pöschl-Teller Potential

M. Hamzavi[1*], A. A. Rajabi[2]

[1]*Department of Basic Sciences, Shahrood Branch, Islamic Azad University, Shahrood, Iran*

[2]*Physics Department, Shahrood University of Technology, Shahrood, Iran*

*Corresponding author: Tel.:+98 273 3395270, fax: +98 273 3395270*

Email: majid.hamzavi@gmail.com

**Abstract**

The trigonometric Pöschl-Teller (PT) potential describes the diatomic molecular vibration. By using the Nikiforov-Uvarov (NU) method, we have obtained the exact analytical s-wave solutions of the radial Schrödinger equation (SE) for the trigonometric PT potential. The energy eigenvalues and corresponding eigenfunctions are calculated in closed forms. Some numerical results are presented too.

## 1. Introduction

The solution of the fundamental dynamical equations is an interesting phenomena in many fields of physics and chemistry. The exact solutions of the SE for a hydrogen atom (Coulombic) and for a harmonic oscillator represent two typical examples in quantum mechanics [1-3]. The Mie-type and pseudoharmonic potentials are also two exactly solvable potentials [4-5]. Many authors exactly solved SE with different potentials and methods [6-16].

The trigonometric PT potential proposed for the first time by Pöschl and Teller [17] in 1933 was to describe the diatomic molecular vibration. Chen [18] and Zhang et al. [19] have studied the relativistic bound state solutions for the trigonometric PT potential and hyperbolical PT (Second PT) potential, respectively. Liu et al. [20] studied the trigonometric PT potential within the framework of the Dirac theory.

The trigonometric PT potential is given by

$$V(r) = \frac{V_1}{\sin^2(\alpha r)} + \frac{V_2}{\cos^2(\alpha r)} \qquad (1)$$

where parameters $V_1$ and $V_2$ describe the property of the potential well, $V_1 \rangle 0$, $V_2 \rangle 0$, while the parameter $\alpha$ is related to the range of this potential [20].

This work is arranged as follows: in Section 2, the NU method with all the necessary formulae used in the calculations is briefly introduced and the parametric generalization NU method is displayed in Appendix A. In Sec. 3 we solve SE and give energy spectra and corresponding wave functions. Some numerical results are given in this section too. Finally, the relevant conclusions are given in section 4.

## 2. NU method

The NU method can be used to solve second order differential equations with an appropriate coordinate transformation $s = s(r)$ [21]

$$\psi_n''(s) + \frac{\tilde{\tau}(s)}{\sigma(s)}\psi_n'(s) + \frac{\tilde{\sigma}(s)}{\sigma^2(s)}\psi_n(s) = 0 \tag{2}$$

where $\sigma(s)$ and $\tilde{\sigma}(s)$ are polynomials, at most of second-degree, and $\tilde{\tau}(s)$ is a first-degree polynomial. To find particular solution Eq. (2) by separating of variables, one deals with the transformation $\psi_n(s) = \phi(s)y_n(s)$. It reduces to an equation of hypergeometric type

$$\sigma(s)y_n''(s) + \tau(s)y_n'(s) + \lambda y_n(s) = 0 \tag{3}$$

$\phi(s)$ is defined as logarithmic derivative

$$\frac{\phi'(s)}{\phi(s)} = \frac{\pi(s)}{\sigma(s)} \tag{4}$$

$y_n(s)$ is the hypergeometric-type function whose polynomials solutions are given by Rodrigues relation

$$y_n(s) = \frac{B_n}{\rho(s)} \frac{d^n}{ds^n}\left[\sigma^n(s)\rho(s)\right] \tag{5}$$

where $B_n$ is the normalization constant and the weight function, $\rho(s)$, must be satisfy the condition [21]

$$\frac{d}{ds}w(s) = \frac{\tau(s)}{\sigma(s)}w(s), \qquad w(s) = \sigma(s)\rho(s) \tag{6}$$

The function $\pi(s)$ and the parameter $\lambda$, required for this method, are defined as follows

$$\pi(s) = \frac{\sigma' - \tilde{\tau}}{2} \pm \sqrt{\left(\frac{\sigma' - \tilde{\tau}}{2}\right)^2 - \tilde{\sigma} + k\sigma} \qquad (7a)$$

$$\lambda = k + \pi'(s) \qquad (7b)$$

In order to find the value of $k$, the expression under the square root must be square of polynomial. Thus, a new eigenvalue equation is

$$\lambda = \lambda_n = -n\tau' - \frac{n(n-1)}{2}\sigma'' \qquad (8)$$

where

$$\tau(s) = \tilde{\tau}(s) + 2\pi(s) \qquad (9)$$

and its derivative must be negative [21]. In this regard, one can derive the parametric generalization version of the NU method [22] where is displayed in Appendix A.

## 3. Solution of radial SE with the trigonometric PT potential

To study any quantum physical system characterized by the empirical potential given in Eq. (1), we solve the original SE which is given in the well known textbooks [1-2]

$$\left(\frac{P^2}{2m} + V(r)\right)\psi(r,\theta,\varphi) = E\psi(r,\theta,\varphi) \qquad (10)$$

where the potential $V(r)$ is taken as trigonometric PT potential form in Eq. (1). Using the separation method with the wave function $\psi(r,\theta,\varphi) = \frac{R(r)}{r}Y_{lm}(\theta,\varphi)$, we obtain the following radial SE as

$$\left[\frac{d^2}{dr^2} + 2m\left(E - \frac{V_1}{\sin^2(\alpha r)} - \frac{V_2}{\cos^2(\alpha r)}\right) - \frac{l(l+1)}{r^2}\right]R_{nl} = 0 \qquad (11)$$

Since the SE with the trigonometric PT potential has no analytical solution for $l \neq 0$ states, we solve s-wave solution of Eq. (11), i.e. $l = 0$. Thus Eq. (11) reduces to the following as

$$\left[\frac{d^2}{dr^2} + \varepsilon - \frac{V_1'}{\sin^2(\alpha r)} - \frac{V_2'}{\cos^2(\alpha r)}\right]R_{n,0} = 0 \qquad (12)$$

where $\varepsilon = 2mE$, $V_1' = 2mV_1$ and $V_2' = 2mV_2$. To solve Eq. (12) by the NU method, we use an appropriate transformation as $s = \sin^2(\alpha r)$ and Eq. (12) reduces to

$$\frac{d^2R_{n,0}(s)}{ds^2} + \frac{\frac{1}{2}-s}{s(1-s)}\frac{dR_{n,0}(s)}{ds} \tag{13}$$
$$+\frac{1}{4\alpha^2 s^2(1-s)^2}\left[\varepsilon s(1-s) - V_1'(1-s) - V_2's\right]R_{n,0}(s) = 0$$

Comparing Eq. (14) and Eq. (A1), we can easily obtain the coefficients $\alpha_i$ ($i=1,2,3$) and analytical expressions $\xi_j$ ($j=1,2,3$) as follows

$$\begin{aligned}
\alpha_1 &= \frac{1}{2}, & \xi_1 &= \frac{\varepsilon}{4\alpha^2} \\
\alpha_2 &= 1, & \xi_2 &= \frac{\varepsilon}{4\alpha^2} + \frac{V_1'}{4\alpha^2} - \frac{V_2'}{4\alpha^2} \\
\alpha_3 &= 1, & \xi_3 &= \frac{V_1'}{4\alpha^2}
\end{aligned} \tag{14}$$

The values of coefficients $\alpha_i$ ($i=4,5,...,13$) are found from relations (A6-A10, A12, A20-A21, A25-A26) of Appendix A. The specific values of the coefficients $\alpha_i$ ($i=1,2,...,13$) together $\xi_j$ ($j=1,2,3$) are displayed in table 1. By using (A17), we can obtain the energy eigenvalues of the s-wave trigonometric PT potential as

$$E_{n,0} = \frac{2\alpha^2}{m}(n+\frac{1}{2})^2 + \frac{\alpha}{2m}(2n+1)(\sqrt{\alpha^2+8mV_1} + \sqrt{\alpha^2+8mV_2}) \tag{15}$$
$$+ \frac{1}{4m}(\sqrt{(\alpha^2+8mV_1)(\alpha^2+8mV_2)} + \alpha^2) + V_1 + V_2$$

Some numerical results are give in table 2. We use the parameters $M = 10 fm^{-1}$, $V_1 = 5 fm^{-1}$, $V_2 = 3 fm^{-1}$ and $\alpha = 1.2, 0.8, 0.4, 0.2, 0.02, 0.002$ [20]. As we see from table 2, when potential range parameter $\alpha$ approaches zero, the energy eigenvalues approaches a constant. From Eq. (15) we find that this constant is $V_1 + V_2 + 2\sqrt{V_1V_2}$, i.e. $\lim_{\alpha \to 0} E_{n,0} = V_1 + V_2 + 2\sqrt{V_1V_2}$.

To find corresponding wave functions, referring to table 1 and relations (A18) and (A22) of appendix A, we find the functions

$$\rho(s) = s^{\sqrt{\frac{1}{4}+\frac{V_2'}{\alpha^2}}}(1-s)^{\sqrt{\frac{1}{4}+\frac{V_1'}{\alpha^2}}} \tag{16}$$

$$\phi(s) = s^{\frac{1}{4}(1+\sqrt{1+\frac{4V_1'}{\alpha^2}})}(1-s)^{\frac{1}{2}(1+\sqrt{\frac{1}{4}+\frac{V_2'}{\alpha^2}})} \tag{17}$$

Hence, relation (A19) gives

$$y_n(s) = P_n^{(\sqrt{\frac{1}{4}+\frac{V_2'}{\alpha^2}},\sqrt{\frac{1}{4}+\frac{V_1'}{\alpha^2}})}(1-2s) \tag{18}$$

By using $R_{n,0}(s) = \phi(s)y_n(s)$, we get the radial wave functions from relation (A24) as

$$R_{n,0}(s) = s^{\frac{1}{4}(1+\sqrt{1+\frac{4V_1'}{\alpha^2}})}(1-s)^{\frac{1}{2}(1+\sqrt{\frac{1}{4}+\frac{V_2'}{\alpha^2}})} P_n^{(\sqrt{\frac{1}{4}+\frac{V_2'}{\alpha^2}},\sqrt{\frac{1}{4}+\frac{V_1'}{\alpha^2}})}(1-2s) \tag{19}$$

or, by substituting $s = \sin^2(\alpha r)$;

$$R_{n,0}(r) = N_{n,0}(\sin(\alpha r))^{\frac{1}{2}+\sqrt{\frac{1}{4}+\frac{2mV_1}{\alpha^2}}}(\cos(\alpha r))^{1+\sqrt{\frac{1}{4}+\frac{2mV_2}{\alpha^2}}} P_n^{(\sqrt{\frac{1}{4}+\frac{2mV_2}{\alpha^2}},\sqrt{\frac{1}{4}+\frac{2mV_1}{\alpha^2}})}(\cos(2\alpha r)) \tag{20}$$

where $N_{n,0}$ is normalization constant. In figure 1, we draw some radial s-wave functions of the trigonometric PT potential, i.e. $\frac{R_{n,0}(r)}{r}$, for $n = 0, 1, 2$ and $3$.

## 4. Conclusions

In this article, we have obtained the bound state solutions of the Schrödinger equation for the trigonometric Pöschl-Teller potential by using the parametric generalization of the Nikiforov-Uvarov method. The energy eigenvalues and corresponding eigenfunctions are obtained by this method. Some numerical results are given in table 2 and it is found that when the potential range parameter $\alpha$ goes to zero, the energy levels approach to a constant.

**Appendix A: Parametric Generalization of the Nikiforov-Uvarov method**

The following equation is a general form of the Schrödinger-like equation written for any potential [22]

$$\left[\frac{d^2}{ds^2}+\frac{\alpha_1-\alpha_2 s}{s(1-\alpha_3 s)}\frac{d}{ds}+\frac{-\xi_1 s^2+\xi_2 s-\xi_3}{\left[s(1-\alpha_3 s)\right]^2}\right]\psi_n(s)=0. \tag{A1}$$

We may solve this as follows. When Eq. (A1) is compared with Eq. (2), we get

$$\tilde{\tau}(s)=\alpha_1-\alpha_2 s, \tag{A2}$$

and

$$\sigma(s)=s(1-\alpha_3 s), \tag{A3}$$

also

$$\tilde{\sigma}(s)=-\xi_1 s^2+\xi_2 s-\xi_3, \tag{A4}$$

Substituting these into Eq. (7a), we find

$$\pi(s)=\alpha_4+\alpha_5 s\pm\left[(\alpha_6-k\alpha_3)s^2+(\alpha_7+k)s+\alpha_8\right]^{\frac{1}{2}}, \tag{A5}$$

where

$$\alpha_4 = \frac{1}{2}(1-\alpha_1), \tag{A6}$$

$$\alpha_5 = \frac{1}{2}(\alpha_2 - 2\alpha_3), \tag{A7}$$

$$\alpha_6 = \alpha_5^2 + \xi_1, \tag{A8}$$

$$\alpha_7 = 2\alpha_4\alpha_5 - \xi_2, \tag{A9}$$

$$\alpha_8 = \alpha_4^2 + \xi_3. \tag{A10}$$

In Eq. (A5), the function under square root must be the square of a polynomial according to the NU method. Thus

$$k_{1,2} = -(\alpha_7 + 2\alpha_3\alpha_8) \pm 2\sqrt{\alpha_8\alpha_9}, \tag{A11}$$

where, we define

$$\alpha_9 = \alpha_3\alpha_7 + \alpha_3^2\alpha_8 + \alpha_6. \tag{A12}$$

For each $k$, the following $\pi$'s are obtained. For

$$k = -(\alpha_7 + 2\alpha_3\alpha_8) - 2\sqrt{\alpha_8\alpha_9}, \tag{A13}$$

$\pi$ becomes:

$$\pi(s) = \alpha_4 + \alpha_5 s - \left[\left(\sqrt{\alpha_9} + \alpha_3\sqrt{\alpha_8}\right)s - \sqrt{\alpha_8}\right]. \tag{A14}$$

For the same $k$, from Eqs. (9), (A2) and (A5)

$$\tau(s) = \alpha_1 + 2\alpha_4 - (\alpha_2 - 2\alpha_5)s - 2\left[\left(\sqrt{\alpha_9} + \alpha_3\sqrt{\alpha_8}\right)s - \sqrt{\alpha_8}\right], \tag{A15}$$

and

$$\tau'(s) = -(\alpha_2 - 2\alpha_5) - 2\left(\sqrt{\alpha_9} + \alpha_3\sqrt{\alpha_8}\right)$$
$$= -2\alpha_3 - 2\left(\sqrt{\alpha_9} + \alpha_3\sqrt{\alpha_8}\right) \langle 0, \tag{A16}$$

are obtained. When Eq. (A2) is used with Eqs. (A15) and (A16), the following equation is derived:

$$\alpha_2 n - (2n+1)\alpha_5 + (2n+1)\left(\sqrt{\alpha_9} + \alpha_3\sqrt{\alpha_8}\right) + n(n+1)\alpha_3$$
$$+ \alpha_7 + 2\alpha_3\alpha_8 + 2\sqrt{\alpha_8\alpha_9} = 0. \tag{A17}$$

This equation gives the energy spectrum of the desired problem. From Eq. (6)

$$\rho(s) = s^{\alpha_{10}-1}(1-\alpha_3 s)^{\frac{\alpha_{11}}{\alpha_3}-\alpha_{10}-1}, \tag{A18}$$

and consequently, after substitution in Eq. (5),

$$y_n(s) = P_n^{(\alpha_{10}-1, \frac{\alpha_{11}}{\alpha_3}-\alpha_{10}-1)}(1-2\alpha_3 s), \tag{A19}$$

where,

$$\alpha_{10} = \alpha_1 + 2\alpha_4 + 2\sqrt{\alpha_8}, \tag{A20}$$

$$\alpha_{11} = \alpha_2 - 2\alpha_5 + 2\left(\sqrt{\alpha_9} + \alpha_3\sqrt{\alpha_8}\right), \tag{A21}$$

and $P_n^{(\alpha,\beta)}$ are Jacobi polynomials. Using Eq. (4)

$$\phi(s) = s^{\alpha_{12}}(1-\alpha_3 s)^{-\alpha_{12}-\frac{\alpha_{13}}{\alpha_3}}, \tag{A22}$$

and the general solution becomes:

$$\psi(s) = \phi(s) y_n(s), \tag{A23}$$

$$\psi(s) = s^{\alpha_{12}}(1-\alpha_3 s)^{-\alpha_{12}-\frac{\alpha_{13}}{\alpha_3}} P_n^{(\alpha_{10}-1,\frac{\alpha_{11}}{\alpha_3}-\alpha_{10}-1)}(1-2\alpha_3 s). \tag{A24}$$

Here, alpha functions are given by:

$$\alpha_{12} = \alpha_4 + \sqrt{\alpha_8}, \tag{A25}$$

and

$$\alpha_{13} = \alpha_5 - \left(\sqrt{\alpha_9} + \alpha_3\sqrt{\alpha_8}\right). \tag{A26}$$

In some problems $\alpha_3 = 0$ [22]. For such problems, when

$$\lim_{\alpha_3 \to 0} P_n^{(\alpha_{10}-1,\frac{\alpha_{11}}{\alpha_3}-\alpha_{10}-1)}(1-\alpha_3)s = L_n^{\alpha_{10}-1}(\alpha_{11}s), \tag{A27}$$

and

$$\lim_{\alpha_3 \to 0}(1-\alpha_3 s)^{-\alpha_{12}-\frac{\alpha_{13}}{\alpha_3}} = e^{\alpha_{13}s}, \tag{A28}$$

the solution given in Eq. (A24) takes the form

$$\psi(s) = s^{\alpha_{12}} e^{\alpha_{13}s} L_n^{\alpha_{10}-1}(\alpha_{11}s). \tag{A29}$$

In some cases, one may need a second solution of Eq. (A1) [22]. In this case, if the same procedure is followed, by using (from Eq. (A11))

$$k = -(\alpha_7 + 2\alpha_3\alpha_8) + 2\sqrt{\alpha_8\alpha_9}, \tag{A30}$$

the solution becomes

$$\psi(s) = s^{\alpha_{12}^*}(1-\alpha_3 s)^{-\alpha_{12}^*-\frac{\alpha_{13}^*}{\alpha_3}} P_n^{(\alpha_{10}^*-1,\frac{\alpha_{11}^*}{\alpha_3}-\alpha_{10}^*-1)}(1-2\alpha_3 s), \tag{A31}$$

and the energy spectrum is

$$\alpha_2 n - (2n-1)\alpha_5 + (2n+1)\left(\sqrt{\alpha_9} + \alpha_3\sqrt{\alpha_8}\right) + n(n-1)\alpha_3$$
$$+ \alpha_7 + 2\alpha_3\alpha_8 - 2\sqrt{\alpha_8\alpha_9} = 0. \tag{A32}$$

Pre-defined $\alpha$ parameters are:

$$\alpha_{10}^* = \alpha_1 + 2\alpha_4 - 2\sqrt{\alpha_8},$$
$$\alpha_{11}^* = \alpha_2 - 2\alpha_5 + 2\left(\sqrt{\alpha_9} - \alpha_3\sqrt{\alpha_8}\right),$$
$$\alpha_{12}^* = \alpha_4 - \sqrt{\alpha_8},$$
$$\alpha_{13}^* = \alpha_5 - \left(\sqrt{\alpha_9} - \alpha_3\sqrt{\alpha_8}\right). \tag{A33}$$

**Table 1.** The specific values for the parametric constants necessary for the energy eigenvalues and eigenfunctions

| constant | Analytic value |
|---|---|
| $\alpha_1$ | $\dfrac{1}{2}$ |
| $\alpha_2$ | $1$ |
| $\alpha_3$ | $1$ |
| $\alpha_4$ | $\dfrac{1}{4}$ |
| $\alpha_5$ | $-\dfrac{1}{2}$ |
| $\alpha_6$ | $\dfrac{1}{4}+\dfrac{\varepsilon}{4\alpha^2}$ |
| $\alpha_7$ | $-\dfrac{1}{4}-\dfrac{\varepsilon}{4\alpha^2}-\dfrac{V_1'}{4\alpha^2}+\dfrac{V_2'}{4\alpha^2}$ |
| $\alpha_8$ | $\dfrac{1}{16}+\dfrac{V_1'}{4\alpha^2}$ |
| $\alpha_9$ | $\dfrac{1}{16}+\dfrac{V_2'}{4\alpha^2}$ |
| $\alpha_{10}$ | $1+\dfrac{1}{2}\sqrt{1+\dfrac{4V_2'}{\alpha^2}}$ |
| $\alpha_{11}$ | $2+\dfrac{1}{2}(\sqrt{1+\dfrac{4V_1'}{\alpha^2}}+\sqrt{1+\dfrac{4V_2'}{\alpha^2}})$ |
| $\alpha_{12}$ | $\dfrac{1}{4}(1+\sqrt{1+\dfrac{4V_1'}{\alpha^2}})$ |
| $\alpha_{13}$ | $-\dfrac{1}{2}-\dfrac{1}{4}(\sqrt{1+\dfrac{4V_1'}{\alpha^2}}+\sqrt{1+\dfrac{4V_2'}{\alpha^2}})$ |
| $\xi_1$ | $\dfrac{\varepsilon}{4\alpha^2}$ |
| $\xi_2$ | $\dfrac{\varepsilon}{4\alpha^2}+\dfrac{V_1'}{4\alpha^2}-\dfrac{V_2'}{4\alpha^2}$ |
| $\xi_3$ | $\dfrac{V_1'}{4\alpha^2}$ |

**Table 2.** The bound state energy levels $E_{n,0}$ for the trigonometric PT potential

| n | $E_{n,0}$ | | | | | |
|---|---|---|---|---|---|---|
| | $M=10fm^{-1}, V_1=5fm^{-1}, V_2=3fm^{-1}$ [20] | | | | | |
| | $\alpha=1.2$ | $\alpha=0.8$ | $\alpha=0.4$ | $\alpha=0.2$ | $\alpha=0.02$ | $\alpha=0.002$ |
| 0 | 18.02560022 | 17.23163309 | 16.47211973 | 16.10494172 | 15.78149898 | 15.74951629 |
| 1 | 22.87051710 | 20.32991862 | 17.95616357 | 16.83082621 | 15.85264289 | 15.75661628 |
| 2 | 28.29143398 | 23.68420415 | 19.50420742 | 17.57271070 | 15.92394680 | 15.76371786 |
| 3 | 34.28835086 | 27.2944896 | 21.11625126 | 18.33059518 | 15.99541071 | 15.77082105 |
| 4 | 40.86126774 | 31 16077522 | 22.79229510 | 19.10447967 | 16.06703463 | 15.77792584 |
| 5 | 48.01018462 | 35.28306074 | 24.53233894 | 19.89436416 | 16.13881854 | 15.78503222 |
| 6 | 55.73510150 | 39.66134628 | 26.33638278 | 20.70024864 | 16.21076245 | 15.79214021 |

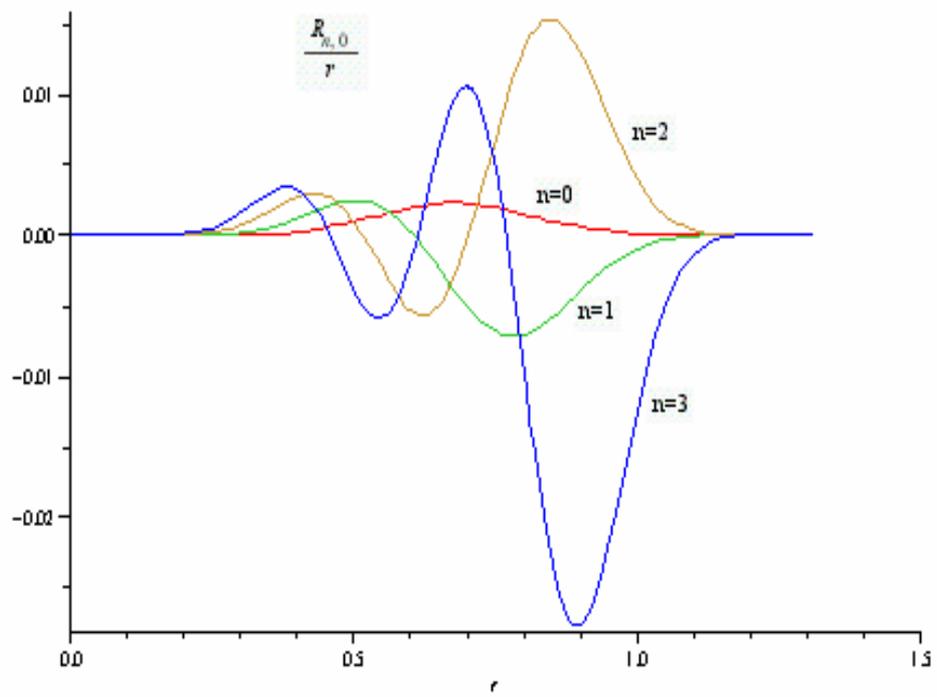

**Figure 1.** The radial wave functions of the trigonometric PT potential